## Ivica Skokić[1]; Roman Brajša[1]

[1] Hvar Observatory, Faculty of Geodesy, University of Zagreb, Kačićeva 26, HR-10000 Zagreb, Croatia

## Abstract

An online software tool for the easy preparation of ephemerides of the solar surface features is presented. It was developed as a helper tool for the preparation of observations of the Sun with the Atacama Large Millimeter/submillimeter Array (ALMA), but it can be used at other observatories as well. The tool features an easy to use point-and-click graphical user interface with the possibility to enter or adjust input parameters, while the result is a table of predicted positions in the celestial equatorial coordinate system, suitable for import into the ALMA Observing Tool software. The tool has been successfully used for the preparation and execution of solar observations with ALMA.

## Keywords:

ALMA, software, solar ephemeris, solar coordinate systems.

## 1. Introduction

The Atacama Large Millimeter/submillimeter Array (ALMA) is a state-of-the-art astronomical radio interferometer designed to observe the Universe in the 35 - 950 GHz frequency range (**ALMA Observatory, 2018**). It is located in Chile at an elevation of 5000 m to avoid most of the water vapour present in the atmosphere, which absorbs radiation in the (sub)millimeter range. Also, the observatory is located away from cities due to radio and light pollution (**Andreić, 2018**). While ALMA has opened a new view into the cold universe, from asteroids to dust inside distant galaxies, it has also enabled exciting new observations of the Sun with high spatial, spectral and temporal resolution (**Bastian et al., 2018**).

During the development of ALMA solar observing modes, there was a need for a user-friendly tool capable of producing ALMA system compatible coordinates for targets on the Sun. For the preparation of ALMA observing proposals and scheduling of observations, ALMA Observing Tool (OT) software is used (**ALMA OT, 2018**). Solar observing modes are very specific compared to other ALMA modes, and this is reflected in the OT, which is lacking some solar specific features. For starters, the OT uses celestial equatorial coordinates for target definition and pointing, while solar physicists use Sun-centered heliocentric and/or heliographic coordinate systems. This makes it complicated for a solar physicist to specify a target on the Sun through the OT. Moreover, the Sun rotates differentially, i.e. the rotation period depends on the latitude, which complicates the process even more. Furthermore, the current OT version still doesn't support retrieval of the latest solar images for visualization and easy definition of the target region, rather the coordinates need to be supplied manually through a text file in the Jet Propulsion Laboratory (JPL) Horizons format. The ALMA Solar Ephemeris Generator (abbr. SEG; **SEG, 2018**) was designed and developed in the Czech ALMA Regional Center Node (**CZ ARC Node, 2018**) with the goal to solve all the above issues and provide a simple and intuitive interface for solar experts.

In the next few chapters, an overview of the SEG user interface is presented, followed by a description of solar and celestial coordinate systems used in SEG, together with the necessary equations to perform the conversion between them. Next, a method for the prediction of the coordinates at some future time is presented and the prediction accuracy is discussed. Finally, specifics of SEG operation and future development steps are described.

## 2. User interface

The SEG software was developed in the HTML/ JavaScript and PHP languages for several reasons. First, it needed to be a multi-platform and easy to use, without complex installations and library dependencies. An application running inside the web browser provides all of this and allows users to run the software without any installation, from a variety of devices and operating systems, even mobile phones and tablets. Moreover, if bugs are found or some last-minute changes are necessary, the

Corresponding author: Ivica Skokić
*ivica.skokic@gmail.com*

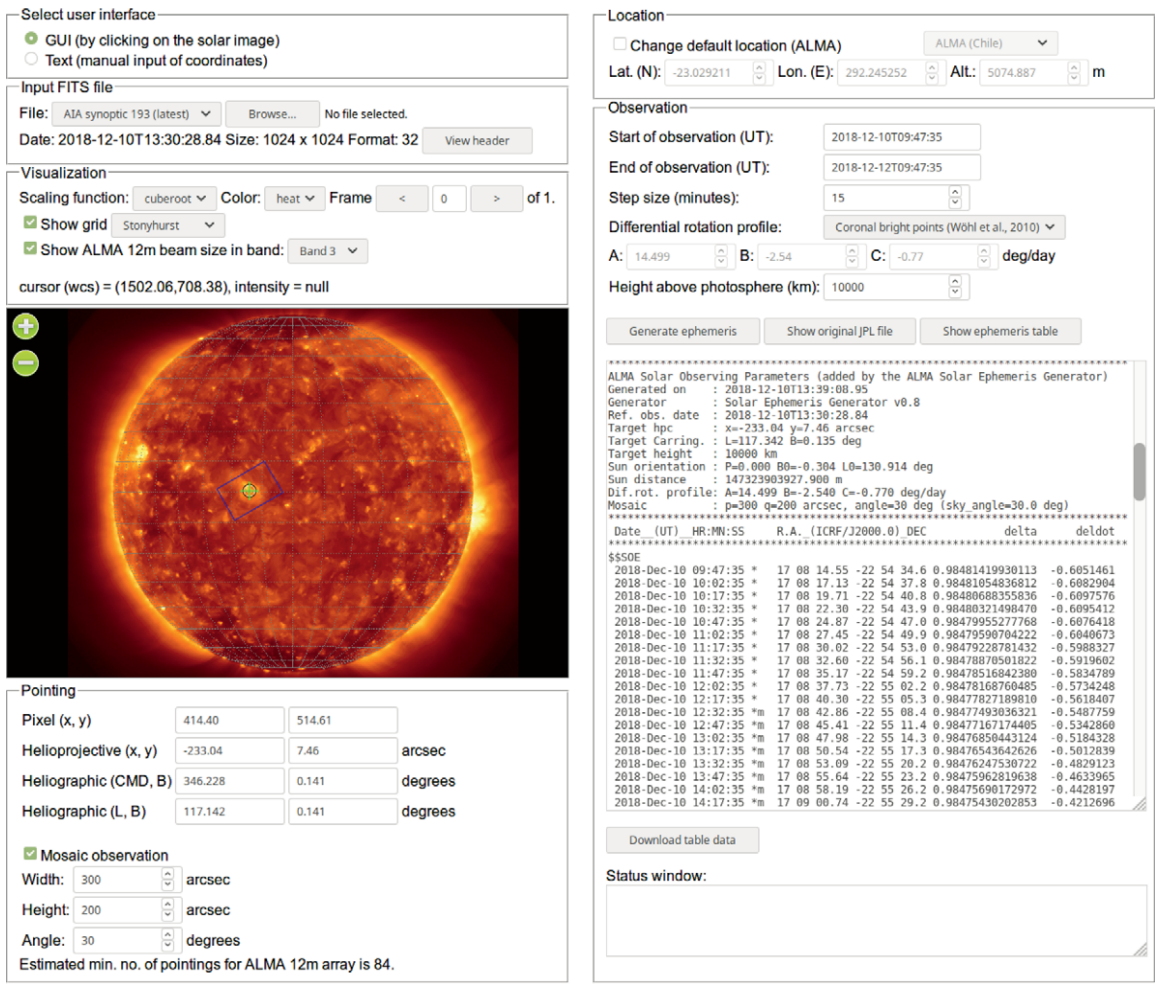

**Figure 1:** The ALMA Solar Ephemeris Generator user interface. Due to document constraints,
the interface is presented here in two images side by side.

author of the tool can implement the changes rapidly and from any computer with an internet connection, regardless of their location. A drawback of the web application approach is that it requires an internet connection.

The user interface is divided into several panels (see **Fig. 1**). In the first panel, the user can decide between a graphical and text user interface. The Graphical User Interface (abbr. GUI) allows the user to set the target by clicking on the desired feature on the latest image of the Sun, while in the text interface, the user has to enter target coordinates manually. In the text interface, "Input" and "Visualization" panels are hidden, while the "Pointing" panel is somewhat different than in the GUI interface. Through the "Input" panel, the user can select an image from the provided list of the latest solar images or upload his/her own. Only the uncompressed Flexible Image Transport System (abbr. FITS) image/data format is currently supported since it provides the necessary metadata within the header, which is used in coordinate transformations. A detailed description of the FITS file format is given by **Pence et al. (2010)**. By default, a list of the Solar Dynamics Observatory/Atmospheric Imaging Assembly (abbr. SDO/AIA; **Lemen et al., 2012**) synoptic images of the Sun in various wavelengths is given. Different solar features are visible in different wavelengths, so the user can easily identify and mark the feature he wants to observe, simply by selecting the right wavelength image. The images are retrieved from the Joint Science Operations Center website (**JSOC, 2018**) in 1/4 of the original resolution (4096x4096 pixels) to save bandwidth and improve loading time. This resolution is still good enough to identify desired features.

Image parameters like the colormap and data scaling function can be tweaked in the "Visualization" panel, for better contrast and the easier identification of solar features. Also, a few optional overlays are available, such as coordinate grids of several different coordinate systems, as well as the ALMA main array 12m antenna beam size in different ALMA frequency bands to give the user a visual estimate of the single antenna field of view. The "Visualization" panel also allows the user to zoom and pan into the region of interest in the image, displaying the current mouse coordinates.

To set the target, the user simply clicks on the desired position in the image. A green cross is displayed at the target position and its coordinates in several solar coor-dinate systems are shown in the table below the image, inside the "Pointing" panel. Here, manual specification or correction of the coordinates is possible in any of the offered coordinate systems. The "Pointing" panel also allows for the visualization of mosaic observations which are used for extended targets when a single pointing field of view is too small. During mosaicing, antennas perform observations of successive fields in an orderly fashion until the whole area is covered. After that, the process starts again and runs in a loop until the end of the observing session. In SEG, it is possible to define mosaic size and orientation which covers the whole area of the region of interest. Note that this angle is in the heliocentric frame (0 degrees means solar east-west). To get the sky angle (in celestial equatorial frame) as required by the ALMA OT, the user should add solar *P* angle to the SEG angle value (**Eq. 1**):

$$angle\_sky\ (OT) = angle\_sun\ (SEG) + P. \quad (1)$$

An estimate of the antenna pointings needed is also displayed and calculated from the **Equations 2** to **5**, given in the ALMA Technical Handbook (**Warmels et al., 2018**):

$$FWHM = 1.13\frac{\lambda}{D} \quad (2)$$

$$N_x = (integer)\left(\frac{L_x}{0.511 FWHM} + 1\right) \quad (3)$$

$$N_y = (integer)\left(\frac{2L_y}{0.511 FWHM \sqrt{3}} + 1\right) \quad (4)$$

$$N_p \approx (2N_x - 1)\frac{N_y}{2} \quad (5)$$

where:

$FWHM$ – the ALMA antenna beam width at half maximum,
$\lambda$ – wavelength,
$D$ – the antenna diameter (12 m for the ALMA main array antenna),
$L_x$ and $L_y$ – the horizontal and vertical angular sizes of the mosaic region,
$N_p$ – the total number of antenna pointings.

Since this is only an estimate, the actual number of pointings as reported by the OT can slightly differ from this one reported by the SEG.

In the "Location" panel, the user can specify the location of the observatory. Since the SEG was developed for the ALMA observatory, a default location is set to ALMA and this panel did not exist in the first software version. However, location specification was added later by request from other observatories which may also use the SEG for preparation of their solar observations. Several predefined locations are listed but the user can also enter the geographic coordinates manually to specify any location on Earth.

The "Observation" panel is used to define the parameters of the observation, namely the start and end time of the observation, step size, differential rotation profile and the estimated height of the target above the solar photosphere. A list of some characteristic rotation profiles is given, but the user can define his/her own, if necessary. After the target is selected and all the parameters have been defined, the user should click the button "Generate ephemeris". This will generate an ephemeris file suitable for import into the ALMA OT. Two additional buttons, "Show original JPL file" and "Show ephemeris table", are provided as a utility for additional checks. The first one will display the original JPL ephemeris file, generated for the solar center (not for the selected target), while the second one will display a table of target coordinates in various coordinate systems in text table format. Generated ephemeris data can be downloaded by clicking the "Download table data". Finally, errors and warnings that occur during the SEG usage are displayed in the "Status" window.

## 3. Solar coordinate systems

Many solar coordinate systems exist for defining positions on the Sun, but the most frequently used types are the helioprojective and heliographic ones (**Thompson, 2006**). The SEG tool accepts coordinates specified in both of them. The heliographic coordinate system is a spherical system similar to geographic coordinates used on Earth, defined with a radial distance *r* from the solar center, latitude *Θ* measured as an angular distance from the solar equator (positive towards the solar north pole), while the longitude can be measured either from the central meridian as seen from the Earth center, called Stonyhurst longitude *Φ*, or from the prime Carrington meridian, in which case we are talking about Carrington longitude, *Φ_C*. The relation between the two can be expressed as (**Eq. 6**):

$$\Phi_C = \Phi + L_0 \quad (6)$$

where:

$L_0$ – the Carrington longitude of the central meridian as seen from the Earth's geocenter.

On the other hand, the helioprojective Cartesian coordinate system is observer-centric, meaning that the coordinates depend on the position of the observer which must be supplied, usually as the Stonyhurst heliographic longitude $\Phi_0$ and latitude $B_0$ of the observer. Helioprojective coordinates, $\theta_x$ and $\theta_y$, sometimes called solar-x and solar-y, represent west and north angular distances of the feature from the solar disk center, as measured by the observer and are usually expressed in arcseconds.

The SEG software implements the following equations for conversion of the coordinates, as given in (**Thompson, 2006**). Conversion from the helioprojective Cartesian ($\theta_x$, $\theta_y$) into heliographic Stonyhurst (*r*, Θ, Φ) coordinates is performed by first converting to inter-

**Table 1:** Differential (sidereal) solar rotation profiles of different features and their estimated heights above the photosphere, as used in the ALMA Solar Ephemeris Generator.

| Feature | A [deg day⁻¹] | B [deg day⁻¹] | C [deg day⁻¹] | Height [km] | Reference |
|---|---|---|---|---|---|
| Sunspot groups | $14.499 \pm 0.005$ | $-2.64 \pm 0.05$ | 0.0 | 0.0 | **Sudar et al. (2014)** |
| Hα filaments | $14.45 \pm 0.15$ | $-0.11 \pm 0.90$ | $-3.69 \pm 0.90$ | 40 000 | **Brajša et al. (1991)** |
| Coronal bright points | $14.499 \pm 0.006$ | $-2.54 \pm 0.06$ | $-0.77 \pm 0.09$ | 10 000 | **Wöhl et al. (2010)** |

mediate heliocentric Cartesian coordinates ($x, y, z$) using **Equations 7** to **11**:

$$q = D_\odot \cos \theta_x \cos \theta_y \qquad (7)$$

$$d = q - \sqrt{q^2 - D_\odot^2 + r^2} \qquad (8)$$

$$x = d\cos\theta_y \sin\theta_x \qquad (9)$$

$$y = d\sin\theta_y \qquad (10)$$

$$z = D_\odot - d\cos\theta_y \cos\theta_x \qquad (11)$$

where:

$D_\odot$ – the Sun-observer distance.

The Stonyhurst coordinates are then found by **Equations 12** to **14**:

$$r = \sqrt{x^2 + y^2 + z^2} \qquad (12)$$

$$sin\Theta = \frac{y\cos B_0 + z\sin B_0}{r} \qquad (13)$$

$$\tan\left(\Phi - \Phi_0\right) = \frac{x}{z\cos B_0 - y\sin B_0} \qquad (14)$$

When converting from heliographic Stonyhurst to helioprojective Cartesian coordinates, again intermediate heliocentric Cartesian coordinates are used, which can be calculated by **Equations 15** to **17**:

$$x = r\cos\Theta sin\left(\Phi - \Phi_0\right) \qquad (15)$$

$$y = r\left(sin\Theta\cos B_0 - \cos\Theta\cos\left(\Phi - \Phi_0\right)\sin B_0\right) \qquad (16)$$

$$z = r\left(sin\Theta sin B_0 + \cos\Theta\cos\left(\Phi - \Phi_0\right)\cos B_0\right) \qquad (17)$$

which are finally converted to helioprojective coordinates (**Equations 18** and **19**):

$$\tan\theta_x = \frac{x}{D_\odot - z} \qquad (18)$$

$$\sin\theta_y = \frac{y}{\sqrt{x^2 + y^2 + \left(D_\odot - z\right)^2}} \qquad (19)$$

The SEG allows either manual specification of target coordinates or by clicking on the desired feature on an image of the Sun. If the latter method is used, transfor-mation from pixel (image) coordinates ($i, j$) into physical (helioprojective) coordinates is needed. This is performed using **Equations 20** and **21**:

$$\theta_x = CDELT1 \cdot \cos\left(CROTA2\right) \cdot i - CDELT2 \cdot \sin\left(CROTA2\right) \cdot j \quad (20)$$

$$\theta_y = CDELT1 \cdot \sin\left(CROTA2\right) \cdot i + CDELT2 \cdot \cos\left(CROTA2\right) \cdot j \quad (21)$$

where:

CDELT1 (CDELT2) – the horizontal (vertical) pixel scale (usually given in arcsec per pixel),

CROTA2 – the image rotation angle in degrees.

These image parameters, and many others, are specified in the FITS image header (**Calabretta & Greisen, 2002; Greisen & Calabretta, 2002**).

## 4. Prediction of position

From the specified target coordinates at some reference time, the SEG calculates target sky position at some future time taking into account the solar differential rotation and movement of the Sun across the sky. Differential rotation $\omega$ is usually represented in the form (**Eq. 22**):

$$\omega(\Theta) = A + B \cos^2 \Theta + C \cos^4 \Theta \qquad (22)$$

where $A$, $B$ and $C$ are rotation profile parameters derived from measurements. A list of rotation profile parameters used in SEG is given in Table 1. Note that these are sidereal rotation parameters, as seen by the observer fixed relative to distant stars. The SEG also offers an option to treat the supplied helioprojective coordinates of the target as constant, which means no rotation profile will be applied. In that case, the target is fixed relative to the solar disk, as seen from the observer on Earth. This is the default option, and the only available option for targets outside the solar disk. It is useful for, e.g., tracking a solar limb or disk center.

If heliographic Carrington latitude and longitude ($\Theta_0$, $\Phi_{C0}$) of the target are given for some reference time $t_0$, its position at some future time $t$ is given by **Equations 23** and **24**:

$$\Phi_C\left(t\right) = \Phi_{C0} + \left[\omega\left(\Theta_0\right) - \omega_C\right]\left(t - t_0\right) \qquad (23)$$

$$\Theta\left(t\right) = \Theta_0 \qquad (24)$$

where:

$\omega_C$ – the Carrington rotation velocity (14.1844 deg day⁻¹).

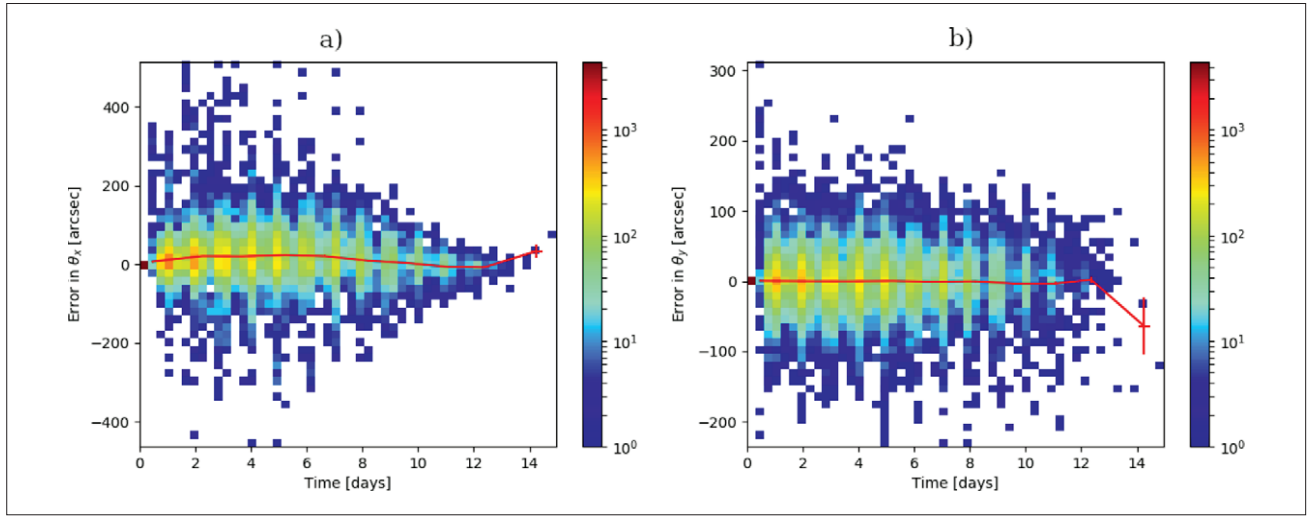

**Figure 2:** Errors of the prediction of coordinates for sunspots in helioprojective *x* coordinate (a) and *y* coordinate (b). Data were binned into 50 bins on each axis. Colors denote number of sunspots in each bin, while the solid red line shows the mean error.

Carrington rotation is necessary in **Equation** 23, because the heliographic Carrington coordinate system rotates with a uniform angular velocity equal to Carrington velocity, referenced to fixed stars. If the Stonyhurst coordinates are used instead, then the synodic rotation profile (as seen from Earth) has to be used, which is a function of time and complicates things significantly, since the Earth's orbital velocity is not constant.

The final step is to calculate the celestial equatorial coordinates $(\alpha, \delta)$ of the target. If $(\alpha_0, \delta_0)$ are the celestial equatorial coordinates of the solar center as seen by the observer, then $(\alpha, \delta)$ can be found from the target's helioprojective Cartesian coordinates $(\theta_x, \theta_y)$ by first finding the angular distance $\rho$ of the target from the solar center and position angle $\varphi$ from the celestial north (**Equations 25** and **26**):

$$\tan\left(\varphi - P\right) = \frac{-\sin\theta_x}{\tan\theta_y} \tag{25}$$

$$\tan\rho = \frac{1}{\cos\theta_x}\sqrt{\sin^2\theta_x + \tan^2\theta_y} \tag{26}$$

where:

$P$ – the position angle of the solar north pole from the celestial north, as seen by the observer and measured eastwards.

Finally, the celestial equatorial coordinates of the target are given by **Equations 27** and **28**:

$$\sin\delta = \sin\delta_0\cos\rho + \cos\delta_0\sin\rho\cos\varphi \tag{27}$$

$$\tan\left(\alpha - \alpha_0\right) = \frac{\sin\rho\sin\varphi}{\cos\rho\cos\delta_0 - \sin\rho\sin\delta_0\cos\varphi} \tag{28}$$

These target coordinates $(\alpha, \delta)$ are then supplied to the ALMA OT.

The question of accuracy of the prediction method described above is of great importance because ALMA currently demands that a solar expert supplies the target coordinates at least two days before the actual observation takes place. The Sun is a highly dynamic object on a wide range of spatial and temporal scales. For instance, the typical lifetime of granules is a few minutes, coronal bright points typically live for several hours, while sunspots and prominences can exist for several days or, in extreme cases, even months. The time and position of emergence of most solar objects is currently impossible to predict. In addition, positions of solar objects are changed by rotation, meridional motions and turbulent convection which makes the prediction of their future position extremely difficult.

For the assessment of accuracy of the above simple prediction method, sunspots seem to be a good candidate because their lifetime spans several days. For this reason, sunspot positions for 2014 from the Debrecen Photoheliographic Data catalog (DPD; **Baranyi et al., 2016; Győri et al., 2017**) were used. Only sunspots with at least two measured positions were selected, resulting in a total of 29413 individual measured positions. Initial sunspot positions were taken as a reference from which future positions were calculated using the rotation profile given by **Sudar et al. (2014)**. Those positions were then compared with the actual measured positions and errors were determined as observed - calculated (O-C) differences, in the helioprojective Cartesian coordinate system. The results are presented in Fig. 2. Errors are shown for each (*x* and *y*) axis separately. The solid red line represents the mean error, while bin colours denote the number of sunspots in each bin. It can be concluded that the mean error of prediction is around zero arcseconds, but individual errors can be as large as 400 and 200 arcsec in horizontal and vertical direction, respec-tively. However, the number of sunspots with such large errors is very small, with most sunspots having errors smaller than 100 and 50 arcsec. The field of view of an ALMA single pointing depends on the frequency, going from 180 arcsec at 35 GHz to 7 arcsec at 950 GHz. This means that for higher frequency bands, the probability of prediction putting the target outside of the field of view is larger, so care must be taken.

## 5. Additional notes

The tool is based on the jsFITS JavaScript library (**js-FITS, 2018**) for FITS file manipulation, with minor modifications. This library supports only uncompressed FITS files which limits available online data sources for synoptic solar image retrieval. Also, the SEG currently only supports the "old" CROTA2 formalism for the specification of image rotations. Support for compressed FITS files and "new" CD/PV matrix rotation formalism is on the roadmap. Background functions for accessing the data from external servers were written in PHP because of the Same-Origin Policy which prevents JavaScript from accessing data from another (external) server. The ephemeris file is generated from the JPL Horizons file retrieved by SEG directly from the JPL Horizons service (**JPL, 2018**). Retrieved coordinates of the Sun are astrometric J2000 coordinates for the ALMA center of array. Atmospheric refraction is not taken into account. Physical ephemeris of the Sun ($P$, $B_0$ and $L_0$) are also from the JPL Horizons. However, during Cycle 4 regular observing sessions, there was a period when the JPL Horizons site went offline and hence the Ephemeris Generator was unable to function properly. Work is currently underway to enable precise ephemeris calculations in the SEG even when the JPL site is not available.

## 6. Conclusion

The Sun is a very specific target for ALMA, so much different than other ALMA objects that it adds additional requirements and constraints on the general ALMA workflow. The ALMA Solar Ephemeris Generator, designed at the CZ ARC node and presented in this paper, is a helper tool developed for easier preparation of the solar observations with ALMA. It was successfully tested during the ALMA solar test campaign in 2015 and regular observing cycles ever since. The SEG provides an easy to use user interface for solar experts to specify targets on the Sun and export the data in a format compatible with the ALMA system. It is continuously improved and the plan is to move the development and code to a public server such as GitHub. Efforts are underway to include the SEG functionality directly into the ALMA Observing Tool. If this turns out to be successful, however, the SEG will not lose its significance, since other observatories and researchers have shown interest in it.

### Acknowledgment

Research leading to this work was performed within ESO Development Plan Study: Solar Research with ALMA (2014 - 2017). This work has been supported in part by the Croatian Science Foundation under the project 7549 "Millimeter and submillimeter observations of the solar chromosphere with ALMA". The authors express their thanks to the ALMA Solar Development Team and Dragan Roša for their comments and suggestions which led to the improvement of the tool.

## SAŽETAK

### Generator efemerida Sunca za ALMA-u

U radu je opisan mrežni program za jednostavnu pripremu efemerida objekata na Suncu. Program je razvijen kao pomoćni alat za pripremu opažanja Sunca pomoću interferometra *Atacama Large Millimeter/submillimeter Array* (ALMA), ali može se koristiti i za pripremu opažanja drugih solarnih opservatorija. Značajke su programa intuitivno korisničko sučelje prilagođeno solarnim ekspertima, unutar kojega je moguće na vrlo jednostavan način definirati objekt i namjestiti parametre opažanja, a kao izlaz program daje tablicu predviđenih položaja objekta u nebeskome koordinatnom sustavu, prilagođeno za unos u sustav ALMA-e preko *Observing Toola*, službenoga alata za pripremu ALMA opažanja. Program je uspješno testiran i korišten za pripremu i izvršavanje opažanja Sunca pomoću interferometra ALMA.

**Ključne riječi:**
ALMA, programska podrška, efemeride Sunca, solarni koordinatni sustavi

## Authors contribution

**Ivica Skokić** (postdoc researcher) designed and developed the software tool presented in this paper and took part in solar ALMA test campaigns. **Roman Brajša** (scientific adviser) initiated this work within the ESO Development Plan Study, provided solar differential rotation profiles and took part in solar ALMA test campaigns.